\documentclass[12pt]{iopart}

\usepackage{iopams}
\usepackage{epsf}
\begin{document}
\title[Nonequilibrium steady states of driven magnetic flux lines]
  {Nonequilibrium steady states of driven magnetic flux lines in disordered 
   type-II superconductors}

\author{Klongcheongsan T$^1$, Bullard T J$^2$ and T\"auber U C$^1$}

\address{$^1$Department of Physics and Center for Stochastic Processes and 
Engineering, Virginia Polytechnic Institute and State University, 
Blacksburg, VA 24061-0435, USA}
\address{$^2$ U.S. Air Force, Wright-Patterson Air Force Base, 
OH 45433-5648, USA}
\ead{\mailto{vmi2004@vt.edu}, \mailto{tauber@vt.edu}}

\begin{abstract}
We investigate driven magnetic flux lines in layered type-II superconductors 
subject to various configurations of strong point or columnar pinning centers 
by means of a three-dimensional elastic line model and Metropolis Monte Carlo 
simulations. 
We characterize the resulting nonequilibrium steady states by means of the 
force-velocity / current-voltage curve, static structure factor, mean vortex 
radius of gyration, number of double-kink and half-loop excitations, and 
velocity / voltage noise spectrum. 
We compare the results for the above observables for randomly distributed point
and columnar defects, and demonstrate that the three-dimensional flux line
structures and their fluctuations lead to a remarkable variety of complex 
phenomena in the steady-state transport properties of bulk superconductors. 
\end{abstract}

\pacs{74.25.Qt, 74.25.Sv, 74.40.+k}
\submitto{\SUST}

\section{Introduction}

From a theoretical and experimental point of view, vortex matter in disordered
high-temperature superconductors has attracted considerable attention during 
the past decades \cite{Blatter}. 
The possibility of practical applications of superconductivity depends on the 
maximum current density which superconductors can carry. 
In type-II superconductors, this is directly related to the flux pinning of 
quantized magnetic flux lines. 
Different experimental techniques such as magnetic decoration \cite{Bolle}, 
scanning Hall probes \cite{Chang}, small angle neutron scattering 
\cite{Gammel}, scanning tunneling microscope imaging \cite{Hess}, and Lorentz 
microscopy \cite{Harada} have been utilized to capture the static structure of
magnetic flux lines pinned by quenched disorder. 
These techniques provide images of the static structure of flux lines on the 
sample's top layer but do not allow mapping out the internal structure of the 
moving flux lines in response to the external current. 
By using neutron diffraction to image the flux lattice \cite{Yaron}, one can 
study the motion of these flux lines, e.g., by compiling a sequence of these 
images of vortex positions into a movie. 

Aside from technological applications of superconductors, the statics and 
dynamics of vortices in type-II materials in the presence of quenched disorder
and external driving force are of fundamental interest also, and have been 
studied quite extensively, both experimentally and theoretically. 
The presence of a small fraction of pinning centers tends to destroy the 
long-range positional order of the Abrikosov flux lattice and results in 
different spatial structures depending on the nature of the material defects. 
In systems with randomly distributed weak point pinning centers, the vortex 
lattice deforms and transforms into a Bragg glass with quasi long-range 
positional order at low fields \cite{Nattermann1}-\cite{Doussal2}.
A vortex glass characterized by complete loss of translational order is 
observed if the pinning strength or magnetic fields are higher 
\cite{Fisher2}-\cite{Fisher3}. 
If correlated defects such as parallel columnar pins are introduced in the 
system, the effective pinning force adds coherently which results in a distinct
strongly pinned Bose glass phase of localized flux lines characterized by 
diverging tilt modulus \cite{Nelson1}-\cite{Nelson2}.
(The term Bose glass stems from a mapping of the statistical mechanics of 
directed lines to bosonic quantum particles propagating in imaginary time 
\cite{Fisher4}, see also Ref.~\cite{Tauber1}.) 
This type of artificial defects can be produced by energetic heavy ion 
radiation \cite{Civale}.

For vortices driven by the Lorentz force induced by external currents, 
analogous moving glass phases have been proposed.
Disorder tends to inhibit flux line motion, and the competition between drive,
pinning, and vortex interactions leads to a variety of different transport 
characteristics.
At low drive, the flux lines remain pinned to the defects.
At sufficiently large driving current, the flux lines will unbind from the
pinning centers and start moving.
At $T = 0$, this depinning transition constitutes a sharp continuous
nonequilibrium phase transition; at finite temperature, this transition is
rounded \cite{Kardar,Fisher5}. 
In the presence of strong randomly distributed point defects, the ensuing 
moving glass is characterized by the decay of translational long-range order, 
the presence of stationary channels of vortex motion, and highly correlated 
channel patterns along the direction transverse to the motion 
\cite{Doussal3,Doussal4}.
For weak point disorder, instead a topologically ordered moving Bragg glass
ensues. 
For intermediate pinning strengths, one expects a moving transverse glass with
smectic ordering in the direction transverse to the flow. 
When correlated disorder is present in the system, the nonequilibrium 
stationary state is predicted to be a moving Bose glass \cite{Doussal4,Olive}.

The principal goal of the present Monte Carlo study is to characterize in 
detail the nonequilibrium steady states of interacting vortex systems in the 
presence of an external driving force and subject to a variety of 
configurations of strong point and/or columnar pinning centers by means of the 
force-velocity / current-voltage curve, the emerging spatial arrangement and 
vortex structures, the corresponding static structure factors, the mean vortex 
radius of gyration, and voltage / velocity noise features. 
This work is complementary to an earlier investigation that utilized the same
model system and numerical algorithm, but addressed the weak pinning regime, 
and specifically focused on the velocity noise spectrum \cite{Bullard2}.
We shall specifically probe the effect of different defect configurations on 
the three-dimensional internal vortex structure that cannot be addressed in
two-dimensional simulations, and is also not easily accessible in experiment. 
Vortex line fluctuations should be expected to be prominent at elevated 
temperatures and near the depinning threshold; randomly placed point defects 
should further enhance thermal line wandering, whereas linear pinning centers
tend to increase the vortex line tension \cite{Blatter}.
Our results allow direct comparison with studies by other authors who implement
a different microscopic scheme, namely Langevin molecular dynamics, to model 
the kinetics of driven three-dimensional vortex systems 
\cite{Olson1}-\cite{Chen2}. 
We remark that it is crucial to explore alternative simulation approaches for
systems that are driven away from equilibrium, in order to ascertain that the
results reflect physical properties rather than algorithmic and modeling
artifacts.
In addition, at low drives we shall explore the flux line creep mechanism via 
thermally activated double-kink configurations, and via vortex half-loops at
intermediate drives \cite{Nelson2}.
Thus we aim to contribute to a better fundamental understanding of electrical 
and transport properties of type-II superconductors subject to various types of
disorder that might aid in further optimization of their desired properties.

\section{Model system and numerical simulation algorithm}

\subsection{Interacting disordered elastic flux line model}

We consider a three-dimensional vortex system in the London limit, where the 
London penetration depth is much larger than the coherence length. 
We model the vortex system by means of an elastic flux line free energy 
described in Ref.~\cite{Nelson1}, see also Refs.~\cite{Dong}-\cite{Petaja}.
The system is composed of $N$ flux lines in a sample of thickness $L$. 
The model free energy $F_N$ (effective coarse-grained Hamiltonian), defined by 
the collection of trajectories of the flux lines labeled by an index $j$ on the
$z$th layer, i.e., two-dimensional vectors $\boldsymbol r_j(z)$, consists of 
three components, namely the elastic energy associated with the line tension, 
the repulsive vortex-vortex interaction potential, and a disorder-induced 
pinning potential \cite{Bullard2}:
\begin{eqnarray}
  F_N(\{ \boldsymbol r_j(z) \}) &=& 
  \frac{\tilde{\epsilon}_1}{2} \sum_{\rm j=1}^N \int_0^L \Bigg\arrowvert 
  \frac{d\boldsymbol r_{\rm j}(z)}{\rmd z} \Bigg\arrowvert^2 \rmd z
  + \frac12 \sum_{\rm i \ne \rm j}^N \int_0^L 
  V\Bigl(|\boldsymbol{r}_{\rm i}(z)-\boldsymbol{r}_{\rm j} (z)|\Bigr) \, \rmd z
  \nonumber \\ &&+ 
  \sum_{\rm j=1}^N \int_0^L V_D\Bigl(\boldsymbol{r}_{\rm j}(z)\Bigr)
  \, \rmd z \ .
\label{contfe}
\end{eqnarray}
Here, the elastic line stiffness or tilt modulus is given in terms of the 
energy scale $\epsilon_0 = \bigl(\phi_0 / 4\pi\lambda_{\rm ab}\bigr)^2$ as 
${\tilde{\epsilon}_1} \approx \Gamma^{-2} {\epsilon_0} \ln (\lambda_{\rm ab}
/ \xi_{\rm ab})$, where $\phi_0 = hc/2e$ is the magnetic flux quantum. 
The parameters $\lambda_{\rm ab}$, $\xi_{\rm ab}$, and $\Gamma^{-2}$ are,
respectively, the in-plane London penetration depth, the coherence length, and 
the effective mass ratio $M_{\bot}/{M_{\rm z}}$, which pertains to a layered 
superconductor model \cite{Lawrence}. 
The expression for the elastic energy \eref{contfe} is valid in the limit 
$\arrowvert \rmd\boldsymbol r_{\rm j}(z) / \rmd z \arrowvert^2 \ll \Gamma^{2}$.
The in-plane repulsive interaction between flux line elements is approximated 
as $V(r) = 2 \epsilon_{\rm 0} K_{\rm 0}(r/\lambda_{\rm ab})$, where $K_0$ 
denotes the modified Bessel function of zeroth order. 
This function diverges logarithmically as $r \rightarrow 0$ and decreases 
exponentially for $r \gg \lambda_{\rm ab}$. 
In our simulations, these vortex interactions are cut off at distance
$L_{\rm y}/2$ in all directions, and the system size is chosen sufficiently 
large in order that numerical artifacts due to this cut-off length are 
minimized. 
We model point and columnar pins through square potential wells of radius $b_0$
with $V_{\rm D}(r) = - \sum_{\rm k=1}^{\rm{N_D}} U_0 \, 
\Theta(b_0 - |\boldsymbol r - \boldsymbol r_{\rm k}^{\rm{(p)}}|\big)$ as our 
coarse-grained defect potential. 
Here, $\Theta$ denotes the Heaviside step function, $N_{\rm D}$ the number of 
effective defect elements, $\boldsymbol r_{\rm k}^{\rm{(p)}}$ the spatial 
coordinates of the $k$th pinning center, and $U_0 \approx \frac{\epsilon_0}{2} 
\ln [1+(b_0/\sqrt{2} \xi_{\rm ab})]^2$ is the interpolated vortex binding 
energy per unit length. 
Finally, we add a phenomenological work term due to the driving force, $W = - 
\sum_{j=1}^N\int_0^L \boldsymbol{f}_{\rm L} \cdot \boldsymbol{r}_{\rm j}(z) \,
\rmd z$. 
In discretized form the model energy \eref{contfe} reads for the $j$th vortex 
line reads (with $z_k = k b_0$)
\begin{eqnarray}
  F_j / b_0 &=& 
  \frac{\tilde{\epsilon}_1}{2} \sum_{k=2}^{L/b_0} \Bigg\arrowvert
  \frac{\boldsymbol r_{\rm j}(z_k)-\boldsymbol r_{\rm j}(z_{k-1})}{b_0} 
  \Bigg\arrowvert^2 + \frac12 \sum_{k=1}^{L/b_0} \sum_{i=1,i \ne j}^{N_{\rm V}}
  V\Bigl(|\boldsymbol{r}_{\rm i}(z_k)-\boldsymbol{r}_{\rm j}(z_k)|\Bigr) 
  \nonumber 
  \\ &&+ \sum_{k=1}^{L/b_0} V_D\Bigl( \boldsymbol{r}_{\rm j}(z_k) \Bigr) - 
  \sum_{k=1}^{L/b_0} \boldsymbol{f}_{\rm L} \cdot \boldsymbol{r}_{\rm j}(z_k)
  \ ,
\label{discfe}
\end{eqnarray}
where $N_{\rm V}$ is the number of flux lines within the radius of the cut-off 
length.

For our simulation, we chose parameter values corresponding to typical material
parameters for YBCO (as listed in appendix D of Ref.~\cite{Nelson2}). 
Throughout this paper, the simulation lengths and energies are reported in 
units of the effective defect radius $b_0$ and interaction energy scale 
$\epsilon_0$ (in cgs units). 
We have set the temperature to $T = 10$ K, and used a pinning center radius and
layer spacing $b_0 = {\rm max}\{c_0,\sqrt{2}{\xi_{\rm ab}}\} = c_0 = 35$ \AA, 
anisotropy $\Gamma^{-1} = 1/5$, average spacing between defects 
$d = 315$ \AA$ = 9.0\, b_0$, and since the temperature is very low, an in-plane
penetration depth $\lambda_{\rm ab} \approx \lambda_0 = 1190$ \AA$ = 34\, b_0$,
and superconducting coherence length 
$\xi_{\rm ab} \approx \xi_0 = 10.5$\AA$ \approx 0.3 \, b_0$. Then
$\epsilon_0 = (\phi_0 / 4\pi \lambda_{\rm ab})^2 \approx 1.9 \times 10^{-6}$
(in cgs units, with dimension energy/length).
The energy scale in the first term of \eref{discfe} is therefore 
${\tilde{\epsilon}_1} \approx 0.18 \, \epsilon_0$, and the pinning strength in
the last term is $U_0 \approx 0.7809 \, \epsilon_0$.
The free energy $F_i$ is measured in units of $\epsilon_0 b_0$.

\subsection{Monte Carlo simulation algorithm}

We employ the standard Metropolis Monte Carlo simulation algorithm in three 
dimensions with the above discretized anisotropic model free energy 
\eref{discfe} \cite{Bullard2,Das}.
The simulations were performed in a system of size 
$[L_{\rm x},L_{\rm y},L]=[\frac{2}{\sqrt{3}} \times 10 \lambda_{\rm ab}, 
10 \lambda_{\rm ab}, 20 b_0]$ with fully periodic boundary conditions at 
temperature $T = 10$ K or $T = 0.002 \, \epsilon_0 b_0$. 
We chose the system's aspect ratio to be $L_{\rm x}/L_{\rm y} = 2/\sqrt{3}$ to
accommodate an even square number of vortices arranged in a triangular lattice 
\cite{Bullard2}. 
We have tested that with a penetration length of $\lambda_{\rm ab} = 35\, b_0$,
the sharp cut-off interaction range of $5 \lambda_{\rm ab}$ in this system size
had no effect on the equilibrium vortex configurations. 
In the absence of a driving force and any pinning sites, square numbers of 
randomly placed vortices were observed to arrange into a six-fold or hexagonal 
lattice after the system equilibrated. 
The average spacing of $9 \, b_0$ between defects gives a total number of 1710 
lines of columnar defects. 
Each columnar defect contains 20 point defect elements, and gives a total of 
1710 $\times$ 20 point defect elements in the system. 
In order to be able to observe any effects due to these pinning centers, the 
maximal displacement for each random move was set to $\Delta = 0.25 \, b_0$. 
This is to guarantee that the flux lines will not move too fast and skip past 
the pinning sites. 
We studied the effect of maximal displacement $\Delta$ on the dynamics of 
vortices and found that if $\Delta$ was too small, the system would be trapped
in metastable states and a much longer simulation time was required to reach 
equilibrium.
We note that the limitation of flux line segment motion induces an artifact in
our simulation \cite{Bullard2}. 
This can be seen in the force-velocity curves in the form of the saturation of 
the curve at high driving force. 
For each move in the simulation, a displacement for the next step is randomly 
chosen from the interval between $-0.25$ and $0.25$.
The acceptance rate for each movement depends on the driving force: the larger 
the driving force, the larger the displacement that each step can take. 
However, the maximal displacement is reached when the driving force is 
increased to a finite value, which results in the saturation of the velocity as
function of drive.
We restricted our explorations to the critical regime where the force-velocity 
curve is not yet saturated.

In this work, we randomly placed the vortex lines in the system at an initial 
high temperature $T = 100$ K and let them equilibrate for 50\,000 Monte Carlo 
steps (MCS) in the absence of any driving force. 
After thus annealing the flux line system, the temperature was suddenly 
quenched to a much lower temperature $T = 10$ K. 
Subsequently, the external drive was applied for another 100\,000 MCS to reach 
a steady state. 
Quantities of interest were then collected every 30 MCS for the next 250\,000 
MCS, and averaged over a number of vortex lines and defect distributions. 
For all types of defect distributions, we tested that flux lines with this
annealing method reached pinned configurations. 
Indeed, annealing yielded the highest critical currents in our simulations. 
In order to facilitate comparison with experimental results, we provide in
table~\ref{dens_table} the magnetic field corresponding to each vortex number 
(16, 36, 64, and 100) used in the simulations, as well as the lattice constants
for the triangular arrays that would result in the absence of disorder.
Velocities will be measured and listed in units of $b_0$ / MCS, and the driving
force (current) will be given in units of $\epsilon_0 / b_0$.
\begin{table}
\begin{center}
\begin{tabular}{c|c|c}
\hline number of vortices & magnetic field $B$ & vortex lattice spacing \\
\hline 16 & $0.019$ T & $101.0 \, b_0$ \\ 
       36 & $0.043$ T & $67.4 \, b_0$ \\
       64 & $0.076$ T & $50.5 \, b_0$ \\
      100 & $0.119$ T & $40.4 \, b_0$ \\ \hline
\end{tabular}
\caption{Magnetic fields and equivalent vortex lattice spacing for the various
   systems with different flux densities used in the simulations. The average
   distance between the 1710 defects in each layer is $9.0 \, b_0$.}
\label{dens_table}
\end{center}
\end{table}

\section{Quantities of interest}

\subsection{Mean velocity}

The force-velocity or current-voltage curve provides important information on 
the non-equilibrium steady state of the driven vortex system.
In experiments, flux lines are subjected to a Lorentz force in the direction 
transverse to the direction of an external current. 
In the simulation, flux lines tend to move in the direction of the driving 
force. 
The mean velocity of flux creep or flux flow is determined by the average of 
the total displacement of each flux line center of mass over a certain number 
of Monte Carlo steps (MCS): 
\begin{eqnarray}
  \boldsymbol{v}_{\rm cm} = \frac{1}{N_{\rm v}} \sum_{i=1}^{N_{\rm v}} 
  \frac{\langle \boldsymbol{r}_{\rm cm, \, i}(t+\tau) 
  - \boldsymbol{r}_{\rm cm, \, i}(t) \rangle}{\tau} \ ,
\end{eqnarray}
where $N_{\rm v}$ is the number of vortex lines, 
$\boldsymbol{r}_{\rm cm , i}$ denotes the center of mass of the $i$th vortex 
line, and $\tau$ is set to 30 MCS. 
Experimentally, the average velocity of the moving vortices is directly related
to the voltage drop across the sample from 
$\boldsymbol{E} = \boldsymbol{B} \times \boldsymbol{v} / c$. 
The driving force arises from the Lorentz force acting on the vortices when 
there is an external current applied to the sample, 
$\boldsymbol{f} = \boldsymbol{J} \times \phi_0 \, \boldsymbol{B} / B$. 
After obtaining the velocity for given drive, we can thus construct the 
force-velocity or current-voltage (I-V) curve and then determine an estimate 
for the critical driving force in each system.

\subsection{Radius of gyration}

\begin{figure}
\centering
\epsfxsize2.3in
\epsffile{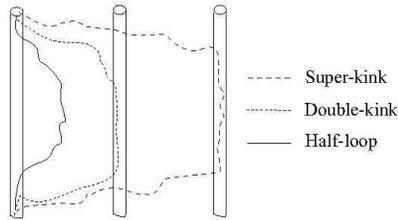}
\caption{Thermally activated half-loop, double-kink, and super-kink 
  excitations.}
\label{figure1}
\end{figure}

The investigation of the three-dimensional structures in nonequilibrium steady
states of moving flux lines is one of the main goals of this work. 
Hence, we are interested in a quantity which reflects the thermal spatial 
fluctuations of the flux lines. 
The mean-square displacement of flux lines measures the total displacement of 
the vortex center of mass from the beginning of the simulation. 
The radius of gyration is the root mean-square displacement of flux line 
elements from their center of mass.  
We use this quantity to directly investigate the effect of the different defect
types on the flux line shapes; in discretized form, its explicit expression in
the direction along the drive is
\begin{eqnarray}
  x_{\rm g} = \biggl( \frac{1}{N_{\rm v} L} \sum_{i , z} \langle 
  [ x_{\rm i}(z) - x_{\rm cm, \, i} ]^2 \rangle \biggr)^{1/2} \ ,
\label{radgyr}
\end{eqnarray}
with $L$ denoting the number of layers (here, $L = 20$). 
Similarly, we can define the mean radius of gyration $y_{\rm g}$ in the
direction transverse to the drive, and perpendicular to the magnetic field.
The total radius of gyration is $r_{\rm g} = \sqrt{x_{\rm g}^2 + y_{\rm g}^2}$.
For systems with columnar defects, we use the maximal displacement in each flux
line to determine the number of half-loop or double-kink excitations, see 
figure~\ref{figure1} \cite{Nelson2}. 
If the maximal displacement in each flux line is approximately equal or greater
than the average distance between defects, this line is considered to have a 
double-kink excitation. 
In contrast, for a maximal displacement greater than the size of defect but 
less than the average distance between defects, we consider this to be a 
half-loop excitation.

\subsection{Static structure factor}

The static structure factor, which is related to the scattering cross section,
reveals spatial symmetries and ordered structures of materials in and out of
equilibrium.
It is basically the Fourier transform of the vortex density-density correlation
function (here, in $d = 2$ dimensions)
\begin{equation}
  I(\boldsymbol q) = \int 
  \rme^{-\rmi \boldsymbol{q} \cdot (\boldsymbol{r}_1 - \boldsymbol{r}_2)} \,
  \langle n(\boldsymbol{r}_1) \, n(\boldsymbol{r}_2) \rangle\, \rmd^\rmd r_1 \,
  \rmd^\rmd r_2 = \langle n(\boldsymbol{q}) \, n(\boldsymbol{-q}) \rangle \ ,
\label{strfac}
\end{equation}
where 
\begin{equation}
  n(\boldsymbol{q}) = \int \rme^{-\rmi \boldsymbol{q} \cdot \boldsymbol{r}} \,
  n(\boldsymbol{r}) \, \rmd^\rmd r 
\end{equation}
is the Fourier transform of the local vortex density $n(\boldsymbol{r})$. 
An ordered or quasi-ordered vortex structure such as the Abrikosov vortex 
lattice or Bragg glass would yield a periodic pattern of sharp peaks in the 
static structure factor. 
On the other hand, a disordered structure such as the vortex glass or vortex 
liquid would result in only a single diffuse peak at $\boldsymbol{q} = 0$.

\subsection{Voltage noise spectrum}

The effect of disorder on the dynamics of driven vortices can also been studied
by means of the velocity or voltage noise power spectrum which is defined by 
\begin{equation}
  S(\omega) = \Bigg| \int \boldsymbol{v}(t) \, \rme^{\rmi \omega t } \, \rmd t
  \Bigg|^2 \ ,
\label{volnoi}
\end{equation}
and directly reflects periodicity of the average velocity in a moving vortex 
lattice. 
The voltage noise spectrum can display broadband or narrowband noise. 
It has been shown that in the presence of weak point defects, and at low 
driving force, vortices are in a plastic flow regime, characterized by a 
broadband noise signal $S(\omega) \sim \omega^{-\alpha}$ \cite{Vestergren}.
This power law may be interpreted as a remnant of the zero-temperature 
continuous depinning transition. 

Narrowband noise is expected to exist provided the moving vortices maintain a 
long-range or quasi long-range positional order as in the moving Abrikosov 
vortex lattice or moving Bragg glass. 
For weak point or columnar pinning centers, a thorough study of the voltage 
noise power spectrum was reported in Ref.~\cite{Bullard2}. 
There, narrowband noise emerged at the characteristic washboard frequency which
is given by $\omega = 2 \pi \langle v \rangle / a$, with the average velocity
$\langle v \rangle$, and the typical intervortex distance $a$. 
The reason for the existence of this washboard noise is that quenched disorder
in the system occasionally traps the vortices.
As a consequence of the periodic vortex structure, this results in a 
periodically varying average overall velocity of the moving Abrikosov lattice 
or Bragg glass. 

The existence of broadband noise and appearance of the washboard frequency in
narrowband noise has been experimentally demonstrated 
\cite{Fiory}-\cite{Togawa} 
and numerically investigated both in two-dimensional 
\cite{Olson2}-\cite{Fangohr2} 
and three-dimensional systems \cite{Bullard2,Chen1}. 
The magnitude of the broadband noise strongly depends on the applied current 
and the external field \cite{Marley}. 
It was shown that the broadband noise in the `peak effect' regime appears with 
the onset of vortex motion. 
The noise power increases to its maximum slightly above the critical current,
and then decreases at large applied currents where the noise becomes 
suppressed.
It was also suggested that the magnitude of the broadband noise would reach 
its maximum in the regime of plastic flow, and in three-dimensional 
simulations, narrowband noise with washboard frequency peaks was reported in 
the moving Bragg glass in the presence of weak pinning disorder 
\cite{Bullard2}.
In a recent experiment, narrowband noise was detected even in the peak effect 
regime, indicating apparent long-range temporal correlations of vortices near 
the critical temperature $T_c$ \cite{Pautrat}. 
The characteristic frequency found in this experiment did not coincide with the
washboard frequency, however: the associated length scale matched the sample 
width rather than the intervortex distance. 
Voltage noise has also been used to study the transition from a moving 
disordered phase, such as the moving vortex glass, to a moving ordered phase,
such as moving Bragg glass or moving Abrikosov vortex lattice 
\cite{Chen1,Olson2}. 
The existence of washboard noise in systems with strong point or columnar 
defects is one of the issues to be addressed in this present study.

\section{Simulation results}

\subsection{I-V characteristics}

\begin{figure}
\centering
\epsfxsize2.5in
\epsffile{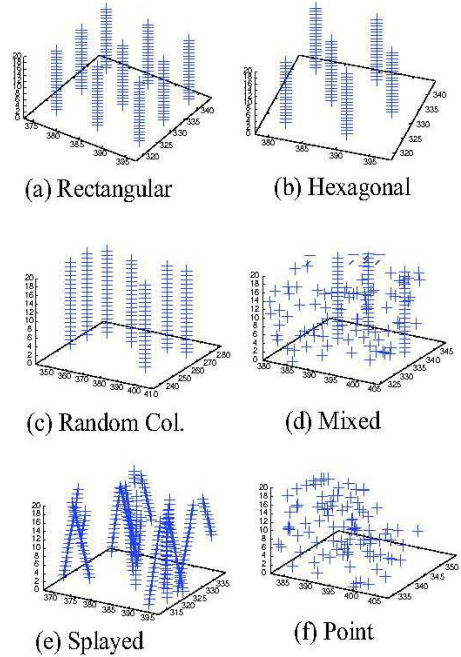}
\caption{Small sample cross sections of systems with (a) rectangularly, (b) 
   hexagonally, and (c) randomly placed columnar defects, (d) a mixed system 
   with both randomly placed point and columnar pinning centers, and systems 
   with (e) randomly placed splayed linear pins, and (f) point defects.}
\label{figure2a_2f}
\end{figure}

We first address the effect of different defect configuration on the I-V 
curves. 
In our simulations, we have studied six defect configurations: regular 
triangular and rectangular arrays of columnar defects; randomly placed splayed
columnar defects; a mixture of point and randomly placed columnar defects; and
randomly placed point defects. 
In order to generate columnar pinning centers, defect elements in each layer 
are arranged on top of each other which results in highly correlated defects 
with the highest critical currents. 
For our system with mixed point and linear defects, we chose the ratio between
point and randomly placed columnar defect elements as 1:1. 
Splayed defects are created by placing columnar defects at random positions and
tilting them slightly at fixed angle in random directions. 
In our simulations we tilt them such that the displacement between the top and
the bottom layer for each defect is about 5 $b_0$. 
(A larger lateral displacement value would result essentially in a randomly 
placed point defect configuration and therefore lower critical currents.) 
With the same defect density, the difference in the value of critical currents
for each I-V curve directly reflects the distinct spatial defect 
configurations.
Small sample slices for each defect configuration type are depicted in 
figure~\ref{figure2a_2f}. 

\begin{figure}
\centering
\epsfxsize2.8in \vskip -0.2in
\epsffile{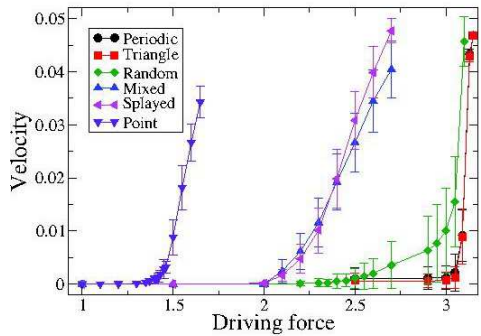} \vskip -0.2in
\caption{Force-velocity or I-V curves for systems with different defect 
   configurations. Each system contains 16 flux lines. The systems with 
   rectangularly (circles) and triangularly (squares) arranged columnar defects
   have higher critical currents than samples with randomly placed columnar 
   defects (diamonds), splayed columnar defects (left triangles), a mixture 
   between points and columnar defects (up triangle), or point defects (down 
   triangle), respectively. This confirms that systems with correlated defects
   yield higher critical currents. Note that here the pinning strength (per 
   unit length) for different defect configuration is set to be the same, 0.78 
   $\epsilon_0 b_0$, while the real pinning strength for point defects is 
   smaller by approximately an order of magnitude. Each of the data points was
   obtained from taking an average over 50 different defect realizations. 
   Driving forces are given in units of $\epsilon_0 / b_0$, and vortex 
   velocities are measured in units $b_0$ / MCS.}
\label{figure3}
\end{figure}
\begin{figure}
\centering \phantom{} \bigskip
\epsfxsize2.5in
\epsffile{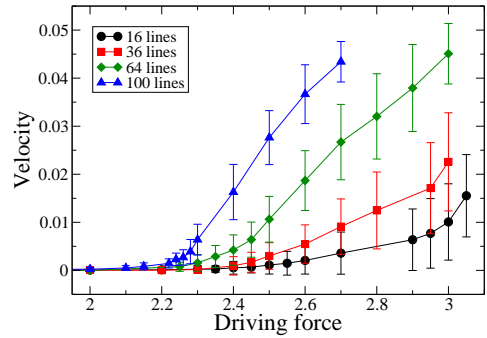}
\caption{Force-velocity or I-V curves for systems with 1710 randomly placed
   columnar defects and different flux densities: 16 lines (circle), 36 lines 
   (square), 64 lines (diamond), and 100 lines (triangle up). The depinning
   force or critical current becomes smaller for denser vortex systems.}
\label{figure4}
\end{figure}
\begin{figure}
\centering 
\epsfxsize2.5in
\epsffile{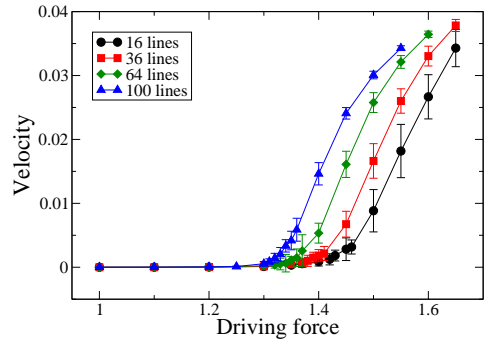}
\caption{Force-velocity or I-V curves for systems with 34200 randomly placed 
   point pinning centers and different flux densities: 16 lines (circle), 
   36 lines (square), 64 lines (diamond), and 100 lines (triangle up). Again, 
   the depinning force or critical current goes down with increasing vortex 
   density.}
\label{figure5}
\end{figure}
\begin{figure}
\centering
\epsfxsize3.0in
\epsffile{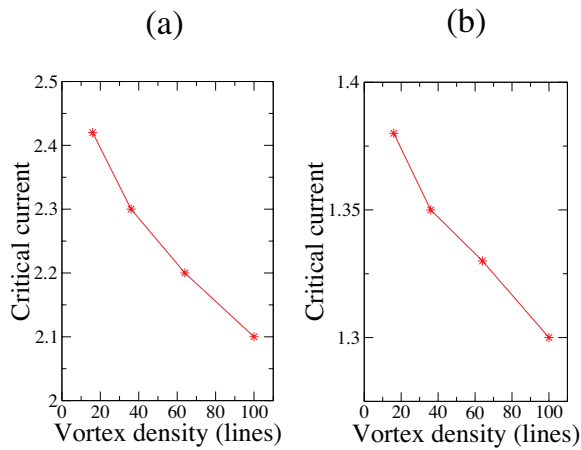}
\caption{Critical current or depinning forces for (a) systems with randomly
   arranged columnar pins and (b) point defects as function of vortex density,
   as obtained from the data shown in figures~\ref{figure4} and \ref{figure5}.}
\label{figure6}
\end{figure}

Characteristic I-V curves for systems with different types of defect 
distributions are shown in figure~\ref{figure3}. 
It is apparent that the critical currents for systems with triangular and 
rectangular arrangement of columnar defects are considerably higher than in
samples with randomly placed columnar pins, randomly placed splayed columnar
defects, a mixture of point and randomly placed columnar pinning centers, and 
randomly placed point defects, respectively. 
This is in good agreement with the experimental and numerical work 
\cite{Blatter,Civale,Bullard2}, which clearly demonstrate that columnar defects
are more effective and thus yield a larger critical current density $J_c$ than 
point pins. 
At low and intermediate drive, flux lines are trapped in metastable states and 
spend most of their time inside or in the immediate vicinity of the pinning 
potential. 
Flux line wandering between nearby defects may occur due to the thermally 
activated jumps. 
The average velocity is extremely small and can be explained by the theory of 
flux creep \cite{Anderson}. 
Far above the critical current, the flux lines flow freely. 
In the case of a system with randomly placed point defects, the pinning force 
density does not add coherently over the length of the vortex. 
Instead, the flux lines will attempt to find energetically optimized paths
through the point pinning landscape, which promotes flux line wandering in the
sample. 
Note that the real pinning strength for a columnar defect is stronger than for 
a point defect by approximately an order of magnitude. 
In our simulations, the pinning strength (per unit length) of all defect 
configurations is set to be the same, comparable to typical columnar pin 
strengths, whereas in Ref.~\cite{Bullard2} a much weaker pinning potential was 
chosen, more in line with typical oxygen vacancy pinning strengths.
We have also confirmed that the depinning current (critical current at $T = 0$)
is lowered for all types of defects in denser vortex systems, where the 
repulsive vortex interactions tend to dominate over the attractive pinning 
energies \cite{Bullard2,Olson1,Kolton1}.
This is demonstrated in figuress~\ref{figure4} and \ref{figure5} for randomly
placed columnar and point defects, respectively, showing the I-V 
characteristics obtained for systems with 16, 36, 64, and 100 lines in either 
case.
In figure~\ref{figure6}, we compare the approximate depinning forces or 
critical currents for systems with varying flux densities and randomly arranged
columnar or point pinning centers, as obtained from the data shown in 
figures~\ref{figure4} and \ref{figure5}.

\subsection{Radius of gyration}

Next we determined the root mean-square displacement or radius of gyration
\eref{radgyr} near the depinning current, which allows us to obtain additional
information about the shape of the moving flux lines in our three-dimensional 
samples. 
(This quantity is obviously not available in a two-dimensional simulation.) 
The radius of gyration represents the spatial fluctuations of the flux line 
from its center of mass. 
We omitted results in the regime of low driving forces since the dynamics there
is extremely slow. 
Data in the regime of large driving forces are discarded also due to the
unphysical limitation for each line element displacement as mentioned in the 
previous section. 

\subsubsection{Columnar defects.}

\begin{figure}
\centering
\epsfxsize2.8in \vskip -0.2in
\epsffile{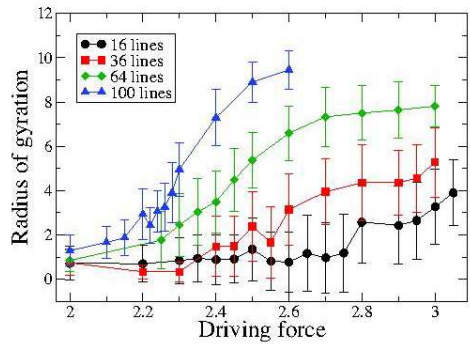} \vskip -0.2in
\caption{Mean radius of gyration (in units of $b_0$) along the drive direction 
  for systems with randomly placed columnar defects with different flux 
  densities: 16 lines (circle), 36 lines (square), 64 lines (diamond), and 100 
  lines (triangle up). The radius of gyration tends to grow at all drives as 
  the density of flux lines increases, but saturates for dense systems. Recall 
  that the mean columnar defect separation is 9 $b_0$.}
\label{figure7}
\end{figure}
\begin{figure}
\centering
\epsfxsize2.8in \vskip -0.2in
\epsffile{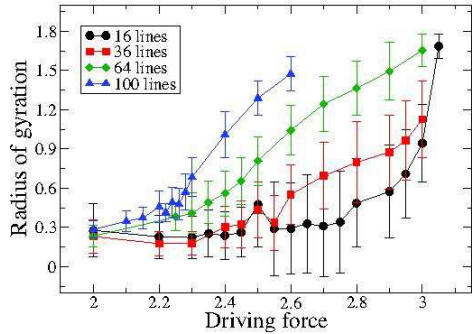} \vskip -0.2in
\caption{Mean radius of gyration in the direction transverse to the drive for 
  systems with randomly placed columnar defects for different flux densities: 
  16 lines (circle), 36 lines (square), 64 lines (diamond), and 100 lines 
  (triangle up).} 
\label{figure8}
\end{figure}
\begin{figure}
\centering
\epsfxsize4.8in \vskip -0.2in
\epsffile{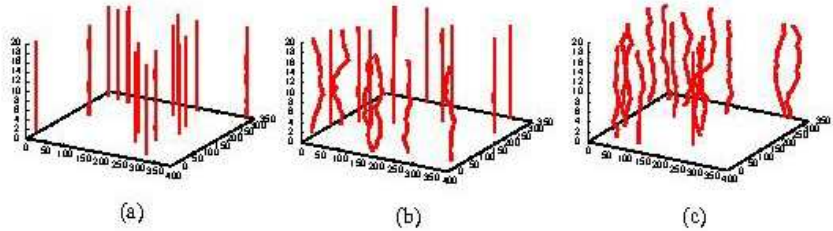} \vskip -0.35in
\caption{Snapshots of (moving) vortices in the presence of randomly placed 
  columnar defects at (a) low, (b) intermediate, and (c) large drive. 
  The flux line system is in the (a) pinned Bose glass phase, (b) partially 
  pinned Bose glass and moving vortex glass / liquid phase, and (c) moving 
  vortex glass / liquid phase.}
\label{figure9abc}
\end{figure}

Figures~\ref{figure7} and \ref{figure8} display our results for the average 
radius of gyration \eref{radgyr} as functions of the driving force in systems 
of 16, 36, 64, and 100 flux lines with 1710 randomly placed columnar defects.
For all investigated vortex densities, the radius of gyration in the direction 
of the drive increases with the driving force in the range studied here. 
At low drives, we observe the Bose glass phase with completely localized vortex
lines inside the columnar defects as shown in the snapshot in 
figure~\ref{figure9abc}a.
In this state, the mean radius of gyration would be smaller than the columnar
pin radius, which is set to 1 in our simulations. 
Since the pinning force of a columnar defect is very large and overcomes the 
repulsive vortex interaction, the flux lines become localized at different 
defect locations resulting in the disordered Bose glass phase. 
At intermediate drives close to the critical regime, some flux lines become 
delocalized and move along the drive direction, with enhanced transverse line
fluctuations. 
This coexistence of a fluid of moving flux lines and the localized Bose glass 
is shown in figure~\ref{figure9abc}b. 
One would expect a small average radius of gyration induced by the moving flux 
lines. 
At high drives, all flux lines are in motion, and hence a larger radius of 
gyration ensues, see figure~\ref{figure9abc}c. 
The radius of gyration also increases with the number of vortices, but 
saturates at large flux densities, in our system at 64 and 100 lines, as shown
in figure~\ref{figure7}.
This increase with flux density is again a consequence of the repulsive vortex
interactions beginning to dominate over the attractive pinning energies.
Flux lines are increasingly caged by their neighbors, and transverse wandering
away from the linear defects becomes a collective excitation affecting several 
vortices. 
The observed saturation indicates that caging due to the effectively stronger 
vortex interaction at large densities governs the flux line dynamics at high 
driving currents. 
Notice that for the largest flux density in our simulation, the gyration radius
reaches beyond the mean columnar pin distance, allowing for pinning of single
lines to several defects.
Qualitatively similar behavior of the mean radius of gyration is seen both
along and transverse to the drive direction; vortex line fluctuations along the
drive direction are however an order magnitude larger than the transverse
fluctuations.

Note that the above results differ characteristically from the data reported in
Ref.~\cite{Bullard2} for systems with much smaller pinning strengths. 
In the moving vortex states of the weak pinning regime, the radius of gyration,
both along the drive and transverse to it, was found to decrease for increasing
flux density. 
This situation corresponds to a dynamics which is dominated by the vortex 
interaction rather than the pinning energies. 
This marked difference to the trend reported for our presented data arises from
the fact that the pinning strength used in the current simulations is stronger 
by a factor of approximately 10: here we address a defect-dominated regime. 
In addition, we simulated our system with a larger penetration length which 
requires a larger system size in order to avoid the numerical artifact due to 
the interaction cut-off. 
With the same number of vortex lines, but in a larger system size and much 
stronger pinning strength, we are here working in a regime where the flux lines
are rather weakly interacting, and their structure and dynamics is dominated 
by the defects. 
Thus, the investigated current range effectively remains rather close to the 
depinning threshold.

\subsubsection{Point defects.}

Point defects directly affect the vortex dynamics. 
In contrast to systems with columnar defects, uncorrelated point disorder 
promotes spatial wandering, transverse to the magnetic field direction, of the 
moving flux lines. 
The spatially randomly distributed pinning centers try to pull the vortex lines
in different directions as they traverse the sample, which leads to less 
effective pinning compared to linear defects. 
In the regime where the defects dominate the dynamics, the radius of gyration 
for point defect is then expected to be larger than for systems with columnar 
pins. 

\begin{figure}
\centering
\epsfxsize2.8in \vskip -0.2in
\epsffile{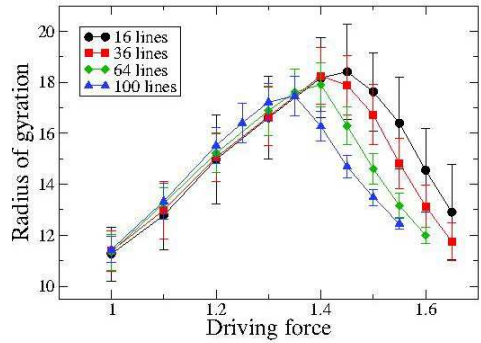} \vskip -0.2in
\caption{Mean radius of gyration along the drive direction for systems with 
  point defects with different flux densities: 16 lines (circle), 36 lines 
  (square), 64 lines (diamond), and 100 lines (triangle up). 
  The radius of gyration starts to grow as the drive increases, reaching its 
  highest value at the critical current; at larger drives, it decreases again.
  Note that even for the largest flux density the peak gyration radius is only 
  about 1/2 the corresponding vortex lattice constant, see 
  table~\ref{dens_table}.}
\label{figure10}
\end{figure}
\begin{figure}
\centering
\epsfxsize2.8in \vskip -0.2in
\epsffile{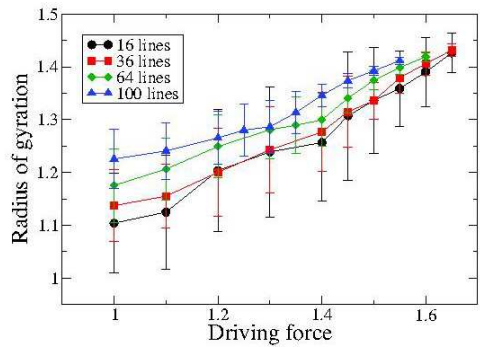} \vskip -0.2in
\caption{Mean radius of gyration transverse to the drive and magnetic field for
  systems with point defects with different flux densities: 16 lines (circle), 
  36 lines (square), 64 lines (diamond), and 100 lines (triangle up). 
  The radius of gyration grows monotonously as the drive increases.}
\label{figure11}
\end{figure}
Figure~\ref{figure10} shows the radius of gyration along the direction of the 
drive. 
The shape of the curve differs from the one depicted in figure~\ref{figure7} 
for randomly placed columnar defects. 
At small drives, the radius of gyration at a specific driving force slightly 
increases with growing flux density, while we observed a larger deviation in 
the system with columnar defects. 
For increasing drives, the radius of gyration keeps growing and reaches its 
maximal value just above the depinning threshold. 
This indicates that the dynamics of the system is only marginally dominated by 
the point pinning centers, and only until the critical threshold is reached. 
In an infinite system at zero temperature, critical fluctuations would lead to
a divergence of the mean radius of gyration at the nonequilibrium depinning
phase transition.
In the moving glass phase, we observe that the gyration radius decreases with
flux density.
At lower densities, point disorder plays a more prominent role, and has the
effect of enhancing flux wandering which allows the vortex lines to more 
efficiently explore favorable sites in the pinning landscape.
We note that the gyration radius along the drive direction peaks at values 
well below the typical vortex lattice constants, see table~\ref{dens_table}.

\begin{figure}
\centering
\epsfxsize4.8in \vskip -0.2in
\epsffile{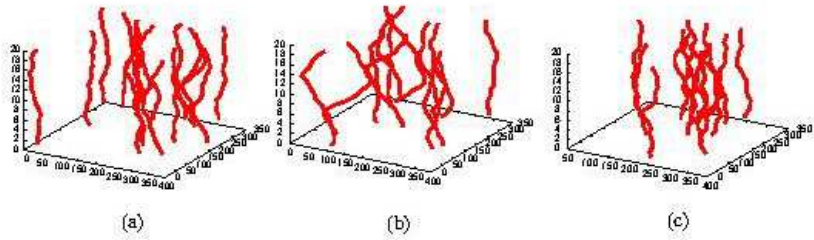} \vskip -0.35in
\caption{Snapshots of moving vortices in the presence of randomly placed point
  defects at (a) low, (b) intermediate, and (c) large drive. 
  Flux lines are in the (a) pinned vortex glass phase, (b) and (c) moving 
  liquid / vortex glass phase. 
  The vortices display the largest average radius of gyration at intermediate 
  drive.}
\label{figure12abc}
\end{figure}

The radius of gyration in the direction transverse to the drive behaves quite
differently. 
As shown in figure~\ref{figure11}, at any vortex density the transverse radius 
of gyration keeps monotonously increasing with the driving force, and no 
tendency of a maximum or saturation is evident. 
The curves tend to merge towards the same value at high drive. 
The transverse radius of gyration is hardly affected by the drive; flux lines 
can only diffuse about their center of mass in the absence of the drive. 
The distance that the flux line element can move away from the center of mass 
is limited by the elastic line tension, which tries to pull the vortex line 
element back towards its center of mass.
For increasing vortex density, the radius of gyration becomes larger since the 
dynamics becomes increasingly dominated by the vortex interactions rather than
the pinning potentials; displacement of a line element of any vortex thus
typically induces motion of nearby elements of other flux lines as well. 
Snapshots of moving vortices subject to point pinning centers at different 
drives are plotted in figure~\ref{figure12abc}; Figure~\ref{figure12abc}b at 
intermediate driving force corresponds to the largest radius of gyration 
observed in our simulations.
Indeed, transverse line wandering is prominent near the depinning current, as
is evident in figure~\ref{figure12abc}b.
(Note, however, that the layer-to-layer displacements still range within the 
applicability limit of the London approximation.)

\subsection{Half-loop and double-kink excitations} 

\begin{figure}
\centering
\epsfxsize2.8in \vskip -0.2in
\epsffile{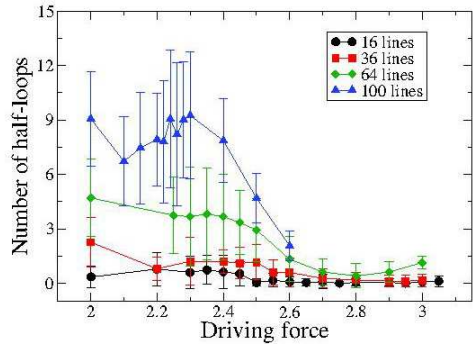} \vskip -0.2in
\caption{Measured number of half-loops for systems with random columnar defects
  with different flux densities: 16 lines (circle), 36 lines (square), 64 lines
  (diamond), and 100 lines (triangle up). 
  The number of half-loops shows a maximum just above the critical current and
  then tends to decrease at all drives as the density of flux lines increases.}
\label{figure13}
\end{figure}
\begin{figure}
\centering 
\epsfxsize2.8in \vskip -0.2in
\epsffile{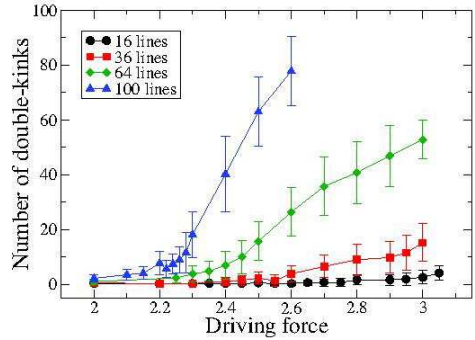} \vskip -0.2in
\caption{Measured number of double-kinks for systems with random columnar 
  defects with different flux densities: 16 lines (circle), 36 lines (square), 
  64 lines (diamond), and 100 lines (triangle up). 
  The number of double-kinks tends to increase with growing vortex density at 
  all drives, similar to the mean radius of gyration.}
\label{figure14}
\end{figure}
The crossover of the dynamics from a disorder- to an interaction-dominated
regime as induced by the driving force has been explained in the previous 
section. 
A large mean vortex radius of gyration appears at elevated driving forces for 
systems with both randomly placed point and columnar defects. 
At low driving force, the structure of the flux line system is different in
both cases. 
Columnar defects tend to straighten the flux lines, whereas in a system with 
point defects flux line wandering is promoted. 
At low driving force, we observed the localized Bose glass in systems with 
columnar defects. 
At low temperatures and low external driving current, flux creep happens via
the formation of vortex half-loop, double-kink, and superkink excitations (see
figure~\ref{figure1}) \cite{Nelson1,Hwa,Tauber2}, which represent saddle point 
configurations whose excitation energies $U_{\rm b}(J)$ represents the 
effective current-depending barrier energy that enters the thermal activation 
factor $\exp[- U_{\rm b}(J) / T]$ for flux flow.
In the localized Bose glass phase, these excitations occurred extremely rarely 
in our simulations, preventing us from obtaining statistically significant 
data.

At intermediate driving forces near the depinning threshold, however, half-loop
excitations become quite prominent, as seen in figure~\ref{figure13}, and at
given vortex density, their typical number decreases with increasing drive. 
Also, a larger number of half-loop excitations is observed as the density is 
increased, which indicates again that the vortex interactions play an important
role for the existence of multiple, `coherent' half-loop excitations. 

We do not observe double kinks at small driving currents; they only appear at 
drives larger than the critical depinning force. 
The presence of double-kinks is largely due to an interplay of pulling on the
vortices and strong pinning forces rather than thermal fluctuations.
As a flux line is moving, some vortex elements will enter columnar pinning 
centers. 
These flux line elements become trapped while other segments are pulled along 
the drive direction. 
The vortex line keeps stretching until the elastic energy overcomes the pinning
energy, whereupon the trapped flux line elements depin from the defect. 
This would result in the emergence of double-kinks, as depicted in 
figure~\ref{figure14}.
These prominent double-kinks are likely largely responsible for the observed 
continuing increase of the mean radius of gyration, see figure~\ref{figure7}.

\subsection{Static structure factor}

\begin{figure}
\centering
\epsfxsize2.0in \vskip -0.2in
\epsffile{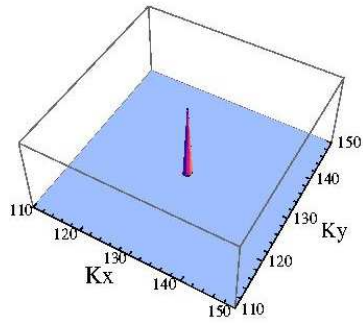}
\caption{Static structure factor plot of moving vortex lines in a system with 
  randomly placed point defects in the nonequilibrium steady state. 
  The strong pinning centers destroy any translational order, and the ensuing
  single structure factor peak indicates a moving plastic or liquid phases of 
  vortices, as also seen in the snapshot \ref{figure12abc}c.
  We obtain similar static structure factors at all vortex densities, driving
  forces, and for all defect configurations studied here.}
\label{figure15}
\end{figure}
In this present study, the abundant point or linear pinning centers are always 
sufficiently strong to destroy any spatially ordered structures in the 
accessible range of driving forces.
In contrast with our earlier investigations of vortex transport in the presence
of much weaker pinning sites \cite{Bullard2}, here we observe neither 
positional nor orientational long-range order.
In this strong pinning situation, flux transport happens via plastic or liquid 
motion at drives larger than the depinning force. 
Correspondingly, we observe a single peak in the static structure factor of the
moving vortex system, for all defect configurations investigated here.
This corresponds to a single peak in the diffraction plot as the experimental
signature for plastic motion \cite{Kolton1,Kolton3}.
As an example, figure~\ref{figure15} shows the structure factor \eref{strfac}
obtained from our simulations for a system with point defects. 
The single peak indicates the presence of an isotropic liquid or a disordered 
solid of moving flux lines, as indeed confirmed by the snapshot depicted in 
figure~\ref{figure12abc}c.
(We have also measured the static structure factor in a different system with 
weak pinning defects, and instead observed hexagonal Bragg peaks corresponding 
to the triangular moving vortex lattice or Bragg glass.)

\subsection{Voltage noise spectrum}

Finally, we report and discuss the measured voltage noise power spectrum 
\eref{volnoi} stemming from the vortex motion in systems with randomly placed 
point and columnar defects. 

\subsubsection{Randomly placed point defects.}

\begin{figure}
\centering
\epsfxsize2.8in \vskip -0.2in
\epsffile{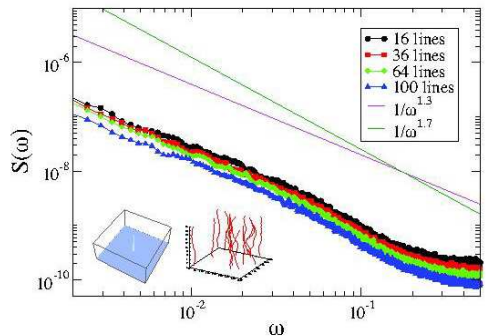} \vskip -0.2in
\caption{Voltage noise along the drive direction for systems with randomly 
  placed point defects at various vortex densities. 
  The driving force along the x-axis for each curve was set to 1.5. 
  The noise amplitude grows as the vortex density increases from 16 to 36, 64,
  and 100 lines in the system. 
  The broadband noise in these systems reflects the presence of a moving 
  disordered phase, such as a moving liquid or plastic phase. 
  This is also supported by the structure factor plot and the snapshot shown in
  the inset. 
  The power laws $S(\omega) \propto 1/\omega^{1.3}$ and 
  $S(\omega) \propto 1/\omega^{1.7}$ are indicated also, to match to the data 
  at low and intermediate frequencies, respectively.}
\label{figure16}
\end{figure}
As shown in figure~\ref{figure16}, we observe broadband noise in our 
simulations of systems with point defects at all vortex densities. 
The results are averaged over 50 defect realizations and initial vortex 
positions. 
We specifically investigate the regime where the driving force is set to 1.5, 
and after the system has reached its nonequilibrium steady state. 
The absence of narrowband noise indicates the presence of a moving vortex 
liquid phase, in agreement with experiments in the peak effect regime 
\cite{Marley}. 
This interpretation is supported by the appearance of the single peak in the 
structure factor and the snapshot of the moving vortices in the inset of 
figure~\ref{figure16}. 
Each flux line in this system subject to strong point disorder moves largely
independently, which implies the loss of any washboard signal in contrast with
the results obtained in samples with very weak point defects \cite{Bullard2}. 
In the presence of weak pinning centers, the vortices arrange themselves into 
the quasi long-range ordered moving Bragg glass, and essentially move with the
same speed.
As this ordered structure traverses the disordered substrate with almost
homogeneous speed $\langle v \rangle$, periodic trapping induces a narrowband 
washboard noise signal with a characteristic peak frequency 
$\omega = 2\pi \langle v \rangle / a_0$, where $a_0$ denotes the vortex lattice
constant.

The magnitude of the broadband noise signal decreases for increasing vortex 
density, indicating that the flux lines tend to reorder at higher flux density.
This is also reflected by the decrease of the radius of gyration for increasing
vortex density. 
The magnitude of the noise indicates the fluctuations of the velocity about its
mean. 
At higher flux densities, the vortex interactions dominate and cause a 
stiffening of the flux lines as they become caged by their neighbors. 
Stronger vortex interactions favor the Abrikosov vortex lattice or Bragg glass 
with reduced noise.
A decrease of the noise power for increasing vortex density or interaction 
strength has also been reported in experimental data \cite{Marley} and 
numerical simulations \cite{Olson2}. 

\begin{figure}
\centering
\epsfxsize2.8in \vskip -0.2in
\epsffile{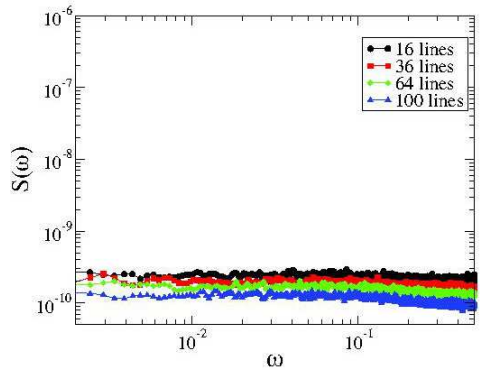} \vskip -0.2in
\caption{Voltage noise in the transverse direction to the driving force 
  (y-direction) for systems with randomly placed point defects at various 
  vortex densities. 
  The driving force along the x-axis for each curve is set to 1.5. 
  Similar to the voltage noise in the x-direction, the magnitude for each curve
  decreases as the vortex density increases from 16 to 36, 64, and 100.  
  The broadband noises in these systems reflect the property of a moving 
  disordered phase.}
\label{figure17}
\end{figure}
A power law broadband noise spectrum $S(\omega) \sim 1/\omega^{\alpha}$ was 
also reported in simulations of two- and three-dimensional XY and dual Coulomb
gas models \cite{Vestergren}, where at zero magnetic field an exponent 
$\alpha=1.5$ was found, whereas $\alpha \approx 1$ in a system with small 
applied magnetic field. 
These results are similar to the experimental data reported in 
Refs.~\cite{Roger,Festin}, where the voltage noise spectrum decays at high 
frequency like $1/\omega$ at elevated temperature and like $1/\omega^{3/2}$ at 
lower temperature. 
The exponent $\alpha=1.5$ also emerges from a functional renormalization group
calculation for point defects to one-loop order in three dimensions 
\cite{Ertas}. 
The voltage noise power spectrum in our data, with strong pinning centers, 
varies approximately like $S(\omega) \sim 1/\omega^{\alpha}$ with 
$\alpha \approx 1.3$, which is well fitted in the low frequency regime, whereas
we find $\alpha \approx 1.7$ at intermediate frequencies. 
Our exponent value is smaller than the exponent $\alpha=2$, which is in fact 
the mean-field value, and was measured in a system of non-interacting driven 
vortices subject to weak point and columnar defects \cite{Das}. 

As shown in figure~\ref{figure17}, the y-component of the broadband noise is 
observed to be nearly flat white noise, and to decrease in magnitude for 
increasing vortex density. 
Transverse to the drive direction, there are much smaller velocity 
fluctuations, and no typical time scale exists.
Each moving flux line can only fluctuate around its center of mass which 
results in zero average velocity.

\subsubsection{Randomly placed columnar defects.}

\begin{figure}
\centering
\epsfxsize2.8in \vskip -0.2in
\epsffile{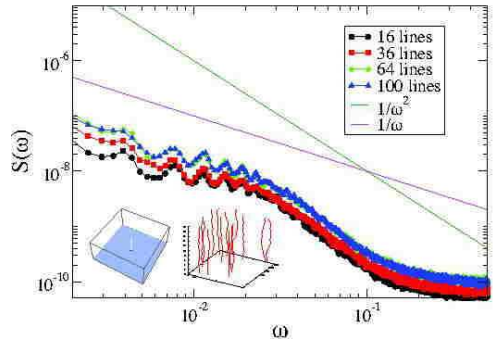} \vskip -0.2in
\caption{Voltage noise power spectrum in the direction of the driving force for
   systems with randomly placed columnar defects at various flux densities. 
   The driving force along the x axis for each curve is set to 2.7. 
   The amplitude for each curve increases with the vortex density.
   The low-frequency oscillations originate from repeated traversal through the
   finite system with periodic boundary conditions.}  
\label{figure18}
\end{figure}

In our system with 1710 randomly placed parallel columnar defects, we observe 
narrowband voltage noise peaks on top of a broadband background, as shown in 
figure~\ref{figure18} for systems with a 16, 36, 64, and 100 flux lines, with 
the data taken in the nonequilibrium steady state at driving force 2.7 along 
the x direction, and the average taken over 50 realizations of the defect and 
initial vortex positions. 
As can be readily confirmed by computing the associated characteristic time,
the presence of these narrowband peaks at low frequencies can however be simply
attributed to the traversal time of the vortices through the entire system.
The observed periodicity is merely due to the moving vortices encountering the
identical linear defect configurations in our system with periodic boundary
conditions.
Even though our system with point defects has similarly strong pinning centers,
no narrowband peaks are observed here since each flux line element moves almost
independently. 
During their repeated traversal through the sample, each flux line element may 
become trapped at point pins at different locations.
The various associated time scales yield quite different frequencies and result
in the characteristic broadband noise.
In comparison to the system with point defects, our sample with randomly placed
columnar pins yields a power law $S(\omega) \sim 1/\omega$ in the low frequency
regime. 
Thus, different defect configurations induce distinct values of the broadband 
noise scaling exponent: correlated defects produce a larger exponent value in
the strong pinning regime.

\section{Discussion and conclusion}

In this paper, we have presented numerical investigations of the nonequilibrium
steady states of driven magnetic flux lines in type-II superconductors subject 
to various configurations of strong point or columnar pinning centers.
We have employed a versatile three-dimensional Metropolis Monte Carlo code 
based on an elastic line model and used the measured current-voltage curve, 
vortex structure factor, mean radius of gyration, number of half-loop and 
double-kink excitations, and the voltage noise power spectrum to thoroughly 
characterize the ensuing moving vortex structures and fluctuations. 

Interesting features arise due to the competition between the elastic energy, 
pinning potential, repulsive vortex interactions, and the driving force. 
In the absence of the driving force and defects, we observed the Abrikosov 
vortex lattice as expected. 
A pinned vortex glass is found in systems with strong point defects, while a 
localized Bose glass is observed in the presence of strong columnar defects. 
From our simulation results, the snapshots of these two systems from the top 
view look very similar, i.e., each flux line looks like a point-like particle 
located inside the pinning center. 
However, these flux line systems behave quite differently if a small driving 
force is applied. 
The effect due to different defect structures is reflected in their different
force-velocity or I-V curves. 
We have found that systems with correlated defects such as parallel or splayed
columnar defects yield higher critical depinning currents. 
The presence of strong point defects strongly promotes flux line wandering, 
which results in a liquid-like vortex structure. 
This feature is also captured by measuring the radius of gyration and studying
snapshots of the driven vortex system. 
By investigating the radius of gyration for various driving forces, we observed
that the driving force changed the moving flux line dynamics from a regime 
dominated by the disorder at low drives to a regime dominated by the vortex 
interactions at high drives. 
This is, for example, evident in the system with point defects as the driving
force or the vortex density are varied. 
In contrast, in systems with weak pinning centers the disorder-dominated regime
is not accessible. 

In the low driving force regime, we did not observe double-kinks excitations,
but a small number of half-loop excitations were detectable at intermediate 
driving forces. 
At large driving force, we observed a liquid-like or amorphous disordered 
structure of moving flux lines in our samples with strong point or columnar 
defects. 
The static structure factor in all our results show a single peak which 
corresponds to a disordered structure. 
Another noticeable difference between the vortex systems subject to point and 
columnar pinning centers emerges when studying the voltage noise spectrum.
While just above the critical threshold broadband noise is observed at all 
vortex densities and driving forces in either case, the characteristic 
exponent describing the power-law decay of the voltage noise signal turns out
to be larger for randomly distributed point defects. 
 
\ack
This material is in part based upon work supported by the U.S. Department of
Energy under Award Number DE-FG02-09ER46613. 
We would like to thank George Daquila, Jayajit Das, and Michel Pleimling for 
helpful discussions.

\end{document}